\documentclass[letterpaper]{sig-alternate-per}
\usepackage[pass]{geometry}

\usepackage{algorithm}
\usepackage{amsfonts}
\usepackage{amsthm}
\usepackage{amsmath}
\usepackage{array}
\usepackage{bbm}
\usepackage{cite}
\usepackage{dsfont}
\usepackage{epsfig}
\usepackage{float}
\usepackage[T1]{fontenc}
\usepackage{graphicx}
\usepackage{indentfirst}
\usepackage{multicol}
\usepackage{subfigure}
\usepackage{url}
\usepackage{multirow}
\usepackage{color}
\usepackage{hyperref}                                           
\hypersetup{
bookmarksnumbered,
pdfstartview={FitH},
linkcolor=blue,
citecolor=blue,
colorlinks=true,
pdfpagemode={UseNone},
plainpages=false
}%


\usepackage{changes}
\definechangesauthor[Arash Asadi]{aa}{orange}
\definechangesauthor[Vincenzo Mancuso]{vm}{red}
\definechangesauthor[Peter Jacko]{pj}{cyan}

\let\mathsetfont\mathcal

\newcommand\setL{\mathsetfont L}

\newcommand\setN{\mathsetfont N}

\newcommand\setS{\mathsetfont S}

\newfont{\mycrnotice}{ptmr8t at 7pt}
\newfont{\myconfname}{ptmri8t at 7pt}
\let\crnotice\mycrnotice%
\let\confname\myconfname%

\begin{document}

\def\sharedaffiliation{%
\end{tabular}
\begin{tabular}{c}}


\permission{The research leading to these results was supported by the CROWD project, under the European Union's Seventh Framework Programme (grant agreement n$^{\circ}$ 318115).\vspace{1mm}}
\title{Modeling Multi-mode D2D Communications in LTE}

\numberofauthors{3}
    \author{
      \alignauthor Arash Asadi\\
      \affaddr{Institute IMDEA Networks}\\
      \affaddr{University Carlos III of Madrid}\\
      \affaddr{Leganes (Madrid), Spain}\\
      \affaddr{arash.asadi@imdea.org}
      \alignauthor Peter Jacko\\
      \affaddr{Lancaster University}\\
      \affaddr{Lancaster, UK}\\
      \affaddr{p.jacko@lancaster.ac.uk}
      \alignauthor Vincenzo Mancuso\\
      \affaddr{Institute IMDEA Networks}\\
      \affaddr{University Carlos III of Madrid}  \\
     \affaddr{Leganes (Madrid), Spain}\\
     \affaddr{vincenzo.mancuso@imdea.org}
                }

%

\maketitle

\begin{abstract}

In this work we propose a roadmap towards the analytical understanding of  
Device-to-Device (D2D) communications in LTE-A networks. Various D2D solutions have been proposed, 
which include {\it inband} and {\it outband} D2D transmission modes, each of which exhibits different pros and cons in terms of complexity, 
interference, and spectral efficiency achieved. We go beyond traditional mode optimization and mode-selection schemes.
Specifically, we formulate a general problem for the joint per-user mode selection, connection activation and resource scheduling of connections.  
\end{abstract}

\vspace{-4mm}
\section{Introduction}
\label{s:intro}
\vspace{-2mm}
The booming growth in popularity of the cellular communications and the exponential rise of cellular data traffic pushed the technology manufactures to their limits in such a way that they could not keep pace with the current demand growth in mobile user's applications~\cite{Bhushan2014Commag}. This made the cellular network industry open to new proposals more than ever. Among various proposals to ameliorate the cellular capacity shortcoming, Device-to-Device (D2D) communication stood out because it detected the paradigm shift in cellular data flow~\cite{asadi2014survey}. The cellular communication ends used to be distant a decade ago, while the emergence of new mobile applications into people's life (e.g., social networking) created significant traffic among nearby users.  The literature on D2D communication is abundant. In fact both academia and industry have been actively exploring use-cases and techniques of D2D communications~\cite{asadi2014survey}.

Academia proposes a wide range of use-cases for D2D communications such as relay~\cite{asadi2014dronee}, multicasting~\cite{zhou_intracluster_2013}, and cellular offloading~\cite{bao_dataspotting_2010}. Initial D2D proposals focused on D2D communication underlaying cellular network transmissions, i.e., using the same spectral resources used for cellular communications~\cite{Doppler2009Commag}. Later, other D2D techniques have been proposed, which either fall under either {\it inband} or {\it outband} D2D communication. Inband D2D communications allow D2D users to communicate over the cellular spectrum, while outband schemes demands the D2D users to access unlicensed bands for D2D transmissions~\cite{asadi2014survey}.  
Each of these D2D operational {\it modes} poses its own merits and disadvantages in terms of interference management, implementation complexity, achievable spectral efficiency, and therefore in terms of performance guarantees. However, the available literature proposes solutions for efficiently implementing each mode in isolation, i.e., {\it mode selection} has not been addressed. Nevertheless, according to the definition provided by 3GPP standards, ``D2D communication is the communication between two users in proximity using a direct link between the devices in order to bypass the eNB(s)\footnote{eNB is the 3GPP term referring to cellular base stations.} or core network''~\cite{lin2013arxiv}. Therefore, any of these modes or perhaps all shall be used for D2D communications. Moreover, promising studies on D2D communication moved industry leaders such as Qualcomm to invest on future implementation of D2D communications, and 3GPP is considering to include generic D2D support in the next release of LTE-A standard as a public safety feature~\cite{lin2013arxiv}.  

In such a framework, we believe that different D2D modes should not be treated as competitors but as complementary techniques. Co-existing D2D modes can immensely increase the system complexity because there should exist a mechanism to select the correct D2D mode according the overall system conditions. 

\vspace{-4mm}
\section{System model}
Our system consists of $ N $ users labelled as $ n \in \setN := \{ 1, 2, \dots, N \} $ in a single-cell LTE network with $20$MHz bandwidth eNB. For notational
consistency, the eNB is labelled as $ N + 1 $. Downlink/uplink channels are open separated bands (i.e., using an FDD scheme). 
Each LTE {\it subframe} ($1$ms) the eNB has 100 time-frequency Resource Blocks (RB)s for downlink and uplink transmission~\cite{johnson2010lte}. 
Users may communicate with other users in the cell or with those outside the cell. 
If a user wants to communicate with another user that is physically close to her, she can use D2D communication. We call such a pair of users a D2D pair. 
We assume that each user wants to communicate only with (at most) one user at any given time.

\textbf{User states.}
The users are allowed to move, and therefore their availability for communication can change over time, so we will say that each user is in a particular
\textit{state} which can change over time. We will denote the state of user $ n \in \setN $ at time $ t $ by $ X_{ n } ( t ) \in \{ 0, 1, 2, \dots, N + 1 \} $,
where each state can be categorized in one of the following types:
\begin{itemize}
\vspace{-2mm}
\itemsep -0.7 mm
\item \textbf{Dormant user (state $ 0 $)}: this is a user who either $ (i) $ has no data to transceive, or $ (ii) $ has a 
poor channel quality in which communication is not feasible.
    
\item \textbf{Cellular user (state $ N + 1 $)}: this is a user who wants to communicate, and can only communicate with the eNB, labelled as $ N + 1 $; 
    
\item \textbf{D2D user (states $ 1 \le m \le N $)}: this is a user who wants and can communicate with her D2D pair labelled $ m $ directly (i.e., she is
    in D2D reach of the user with whom she wants to communicate). 
\vspace{-2mm}
\end{itemize}
Consequently, the number of users in each state will vary in time. However, we assume that state changes occur 
(or are detected by the mode selection mechanism) at regular {\it mode intervals} of duration $T$ seconds.
We denote by $\setS_m (j)$ the set of users in state $m \in \{0, .. , N+1\}$ in mode interval $j$.
Each cellular user and D2D pair is associated with a flow and each pair can only have one active flow at any given time. Moreover, the D2D communication is
assumed to be symmetric, i.e., if user $ n \in \setN $ is in state $ m \in \setN $, then user $ m $ is in state $ n $.



\begin{figure} [!t]
\centering
  \includegraphics[scale=0.40]{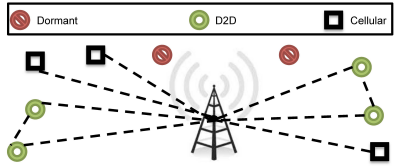}
\vspace{-3mm}
\caption{An illustration of a cell with dormant, cellular, and D2D users}
\label{fig:network}
\vspace{-5mm}
\end{figure}

\textbf{Graph model.}
We can map the network with $ N $ users and one eNB to a graph with $ N + 1 $ nodes, where nodes $ 1 $ to $ N $ represent the users and node $ N + 1 $
represents the base station. The location of nodes in the graph does not necessarily correspond to a physical position of the users (which are moreover allowed
to move within a cell). The users' physical location and mobility affect the arcs of the graph rather than the nodes. An arc between two nodes represents
the communication feasibility between the two nodes. Thus, at every given time, there is an arc between two nodes if these two nodes want to communicate and
their physical channel allows a non-zero transmission rate. Thus, dormant users are isolated (without any arc), cellular users have an arc with eNB, and D2D
users have an arc with their pairs and with eNB. In particular, if a user $ n $ is in state $ m \in \setN $, then there is an arc between users $ n $ and $ m $
and another one between user $ n $ and eNB $ N + 1 $; if a user $ n $ is in state $ N + 1 $, then there is an arc between user $ n $ and eNB $ N + 1 $. Thus,
the state of the user indicates her neighbour(s). See \autoref{fig:network} for an illustration. Due to users mobility and communication needs, which affect
users' states, the arcs will change over time (which is fully captured by state changes each $T$ seconds).

Note that there are at most $ 3N/2 $ arcs in the graph, because each cellular user creates $ 1 $ arc and each D2D pair create $ 3 $ arcs. The arcs will be
denoted by their end-nodes, $ ( n, m ) $. We will further denote the existence of arc $ ( n, m ) $ at time $ t $ by $ Z_{ n, m } ( t ) \in \{ 0, 1 \} $.


\textbf{Cellular mode.} Users in state $N + 1$ use normal cellular communication. We define this as mode $ 0 $.

\textbf{D2D modes.}
Every D2D pair can communicate via any of the following modes (see \autoref{fig:in-outband}): 
\begin{itemize}
\vspace{-2mm}
\itemsep -0.7 mm
\item \emph{Underlay inband} (mode $1$): D2D users reuse the RBs which are available to the cellular users (and therefore share resources with connections in mode $0$).
\item \emph{Overlay inband} (mode $2$): D2D communications occur over dedicated RBs,  subtracted from cellular users. 
\item \emph{Outband} (mode $3$): D2D users switch to WiFi.
\vspace{-2mm}
\end{itemize}
In both underlay and overlay modes, D2D pairs can use the same RBs used by other D2D pairs simultaneously  as long as interference allows. \autoref{tb:adv-disadv} summarizes the merits and drawbacks of each method. Note that the major issue in inband is interference control, while outband D2D suffers from the power consumption of WiFi interface.

\begin{figure} [!t]
\centering
\includegraphics[scale=0.23]{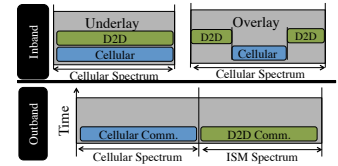}
\vspace{-3mm}
\caption{Schematic representation of overlay inband, underlay inband, and outband D2D.}
\label{fig:in-outband}
\vspace{-5mm}
\end{figure}

\begin{table}[t!]
\scriptsize
\centering
\caption{Cons and pros of each D2D mode}
\label{tb:adv-disadv}
\resizebox{1\columnwidth}{!}{
\begin{tabular}{|p{3.5cm}|cccc|}
\cline{2-5}
\cline{2-5}
\multicolumn{1}{c|}{}							&Underlay					&Overlay					&Cellular					&WiFi	\\	
\hline
\hline
Interference between D2D and cellular users 		&$\checkmark$				&$\times$					&$\times$					&$\times$			\\
\hline
Interference among D2D users		 			&$\checkmark$				&$\checkmark$				&$\times$					&$\times$			\\
\hline
Requires dedicated resources for D2D users 		&$\times$					&$\checkmark$				&$\times$					&$\times$			\\
\hline
Controlled interference environment 				&$\checkmark$				&$\checkmark$				&$\checkmark$				&$\times$			\\
\hline
Simultaneous D2D and cellular transmission		&$\times$					&$\times$					&$\times$					&$\checkmark$		\\
\hline
Increased spectral efficiency					&$\checkmark$				&$\checkmark$				&$\times$					&$\checkmark$		\\
\hline
Energy cost								&Eq.\eqref{eq:einband}		&Eq.\eqref{eq:einband}		&Eq.\eqref{eq:einband}		&Eq.\eqref{eq:pWifi}			\\
\hline
\end{tabular}
}
\vspace{-5mm}
\end{table}	
 




\vspace{-2mm}
\subsection{Joint scheduling and mode selection}

At a given time, every existing arc represents a possible data transmission, and can be either active (allowed to transmit) or inactive (not allowed to
transmit). We have to design a mechanism that selects the arcs to be used in each mode interval, and assign RBs to the arcs. 

There are three tiers of decision making in our system:
\begin{itemize}
\vspace{-2mm}
\itemsep -0.7 mm
\item \textbf{Mode selection}: we have to decide about the operating mode for D2D pairs (modes $1$ to $3$);

\item \textbf{Connection activation}: we have to decide which connections (arcs) are active given the interference constraints of the selected mode;

\item \textbf{Connection scheduling}: we have to decide which connections transmit at what transmission rate (i.e., how the RBs are allocated).
\vspace{-2mm}
\end{itemize}

The three tiers are intertwined, since the interference depends on mode selection but cannot be known before connection activation and scheduling. In turn, connection activation and scheduling depend on which connection is active and on which mode is used in each connection. For sake of tractability, we implement first mode selection, assuming a worst-case interference scenario for activating connections, and then we implement a conventional opportunistic cellular scheduler, Proportional Fair (PF) for connection scheduling at eNB. Specifically, scheduling priorities are computed on 
instantaneous or expected instantaneous channel quality of the users, and RBs allocated to inband overlay are fixed (they are used by users in mode $2$, or released for modes $0$ and $1$ if no connection selects mode $2$). 
In each {\it subframe} (lasting $1$ ms), only one user is scheduled for direct communication to the eNB, while the number of concurrent D2D transmissions is not limited {\it a priori}. Therefore, mode $0$ users do not interfere with each other, mode $1$ users interfere with users in modes $0$ and $1$, and mode $2$ only causes interference among users in 
mode $2$.  

Our system operates in discrete time units and there is a central \textit{controller}, who schedules all the transmissions. 
For tractability, we build the model hierarchically. The controller observes the actual CQI (the LTE Channel Quality Indicator, which corresponds to a particular transmission rate) 
of each connection and takes the fundamental scheduling decisions every \textit{frame} (consisting of 10 subframes, hence lasting $10$ ms). All scheduled transmissions in 
each subframe occur simultaneously and use the maximal transmission rate permitted by the CQI observed by the eNB. A connection scheduled in a subframe will use all the RBs assigned to the specific mode selected. 

The controller further estimates at the beginning of every \textit{mode interval} $T$ the future CQI of all possible connections (both
WiFi and cellular), and decides upon the mode for the duration of the mode interval, which may imply setting up new connections or closing existing ones,
i.e., changes the arcs of the graph. From the graphical point of view, there is hence a new random graph (on a fixed number of nodes $ N $) at the beginning of
every mode interval $T$. The controller also decides which of the arcs are active (allowed to transmit) over the mode interval. In practice, the random
graphs will be strongly correlated, because the mode interval length has to be short enough (say, $200$ frames) to prevent users to move and experience deep channel fading. 


\vspace{-2mm}
\subsection{CQI and interference estimation}

As mentioned above, CQI information is needed for each connection. We assume that the eNB can estimate the CQI of each connection by using the reports produced by the users, containing the signal strength they receive from each and all their neighboring transmitters. By extending this legacy LTE scheme, the interference can be estimated as well.  Thus, the eNB can build the interference table, whose elements $I_{n,m}(j)\ge0$ represent the interference caused by user $ n $ to user $m$ ($\forall n,m \in \setN \cup \{N+1\}$), in mode interval $j$.  Hence, decisions upon connections (set up/close) can be made on the observations from previous mode intervals.

\vspace{-4mm}
\section{Problem Formulation}

We solve the problem hierarchically at the beginning of each mode interval $ j $, i.e., each $T$ seconds. Let $ \setL ( j ) $ be the set of all existing arcs
during mode interval $ j $, i.e., such that $ Z_{ n , m } ( j ) = 1 $.
%
For an active arc $ ( n , m ) $ under an LTE mode $ i \in \{0, 1, 2 \} $ in mode interval $ j $ we define the energy consumption $ E^{ i }_{ n , m } ( j ) $
and the transferred data $ \theta^{ i }_{ n , m } ( j ) $ (both per mode interval $T$) as follows:
\vspace{-2mm}
\begin{eqnarray}
E^{ i }_{ n , m } ( j ) & =&   \left( p^{ i, \text{TX} }_{ n } + p^{ i, \text{RX} }_{ m } \right) B^{ i }_{ n , m } ( j ), \label{eq:einband} \\
\theta^{ i }_{ n , m } ( j ) & =&  B^{ i }_{ n , m } ( j ) R^{ i, \text{CQI} }_{ n , m } ( j ), \label{eq:tinband2}
\end{eqnarray}
where 
we do not consider the baseline energy consumed by a user in LTE in one mode interval, since it cannot be changed unless the node is switched off, 
$ p^{ i, \text{TX} }_{ n } $ and $ p^{ i, \text{RX} }_{ m } $ are the energy consumed by user $ m $ per transmitted and received RB, respectively, $ B^{ i }_{ n , m } ( j ) $ is the number of RBs allocated to arc $ ( n , m ) $, and $ R^{ i, \text{CQI} }_{ n , m } ( j ) $ is the
number of transmitted bits per RB of arc $ (n , m) $ under mode $ i $ during mode interval $ j $. 

For an active arc $ ( n , m ) $ under mode $ 3 $ (i.e.,  WiFi) in mode interval $ j $ we define the energy consumption $ E^{ 3 }_{ n , m } ( j ) $
and the throughput $ \theta^{ 3 }_{ n , m } ( j ) $ (both per mode interval) as follows:
\vspace{-1mm}
\begin{eqnarray}
\label{eq:pWifi}
E^{ 3 }_{ n , m } ( j ) &=& 2 \beta^{ \texttt{WiFi} } + \left( p^{ 3, \text{TX} }_{ n } + p^{ 3, \text{RX} }_{ m } \right) \theta^{ 3 }_{ n , m } ( j ), \\
\theta^{ 3 }_{ n , m } ( j ) &=& T \cdot R^{ i, \text{CQI} }_{ n , m } ( j ),
\end{eqnarray}
where $ \beta^{ \texttt{WiFi} } $ is the baseline WiFi energy consumed by a user in one mode interval, and $R^{ i, \text{CQI} }_{ n , m }$ is the WiFi rate. Note that the energy consumption as defined here can incorporate both the consumption
due to transmission/reception and packet processing (see \cite{asadi2014dronee}).

The utility function for an active arc $ ( n , m ) $ under mode $ i $ in mode interval $ j $ is defined as follows:
\begin{equation}
\label{eq:u}
U_{ n , m }^i ( j ) = \theta^{ i }_{ n , m } ( j ) - \alpha  E^{ i }_{ n , m } ( j ),
\end{equation}
where $ \alpha $ is a relative cost of energy. 
We use a set of binary decision variables $\{Y^i_{n,m}(j)\}$, 
to formulate the problem of mode selection for mode interval $j$, preceding the RB allocation procedure in the above described system (note that at mode selection time 
it is not yet possible to predict the exact interference caused by/to D2D users, so we account for the worst-case interference). The problem is formulated as follows (we omit the dependency on $ j $ from utilities, interferences, and decision variables):
\vspace{-2mm}
\begin{align*}
\text{maximize} &  \sum_{i=0}^{3} \sum_{(n, m)\in\setL( j )}  U_{ n, m }^{ i } Y_{ n, m }^{ i }; \\
\text{s.t.:} \quad	& \sum_{i=0}^{3} \sum_{n|(n, m)\in \setL(j)} Y_{ n , m }^{ i } \le 1  \quad \forall m \in \setN; \\
		& \sum_{i=0}^{3} \sum_{m|(n, m)\in \setL(j)} Y_{ n , m }^{ i } \le 1 \quad \forall n \in \setN; \\
		& \sum_{\substack {(n, m)\in \setL(j) }} Y_{ n , m }^{ 1 } I_{ n,x} \le \gamma \quad \forall x\in \setS_{N+1} \cup \{N+1\};\\
		& \sum_{i \in \{0, 1\}} \sum_{(x, y) \in \setL(j) \setminus \{ (n, m) \} } Y_{ x , y }^{ i } Y_{n, m}^ {1} I_{ x, m} \le \gamma \quad \forall (n, m) \in \setL(j);\\
		& \sum_{(x, y) \in \setL(j) \setminus \{ (n, m) \} } Y_{ x , y }^{ 2 } Y_{n, m}^ {2} I_{ x, m} \le \gamma \quad \forall (n, m) \in \setL(j);
\end{align*}
The formulated problem maximizes the sum of utilities over all possible combinations of users and modes. 
The first and second constraints ensure that at most one active connection can be allowed for each user (but for the eNB, which is labeled as $N+1$). 
The third constraint imposes that the interference caused by inband underlay D2D users to cellular users and to the eNB is below a threshold $\gamma$.
The fourth constraint ensures that the interference caused by cellular and inband underlay transmissions to other inband underlay users is below a threshold. 
Finally, the fifth constraint ensures that the interference caused by inband overlay transmissions (mode $2$) is below the threshold $\gamma$.  
The challenge to be tackled in future work consists in plugging the resource allocation scheme into the computation of $\theta^{ i }_{ n , m }$ and $E^{ i }_{ n , m }$, which, in turn, depend on mode selection and connection activation decisions through the resource allocation scheme. 

\vspace{-4mm}

%


%
%

\bibliographystyle{abbrv}

\renewcommand{\baselinestretch}{0.96}

\scriptsize
\bibliography{biblio}

\end{document}